\documentclass[aps,prb,twocolumn,showpacs,a4paper]{revtex4}

\usepackage{graphicx}

\begin{document} 
\title{ Equilibrium spin currents in the Rashba medium}  
\author{ E.B. Sonin}

\affiliation{ Racah Institute of Physics, Hebrew University of
Jerusalem, Jerusalem 91904, Israel} 

\date{\today} 

\begin{abstract}
We analyze equilibrium spin currents in a 2D electron gas with Rashba spin-orbit interaction (Rashba medium). In a uniform Rashba medium these currents are constant, and their ability to really transport spin is not evident. But even weak inhomogeneity of the Rashba medium allows to reveal that  equilibrium spin currents can transport spin from areas where spin is produced to areas where spin is absorbed.
\end{abstract} 


\maketitle

             
\section{Introduction}

Producing {\em spin currents} is an essential part of spintronics \cite{ZFS}. This explains a growing interest to spin currents in today literature. At the same time the concept of spin current is under intensive theoretical scrutiny and remains to be a controversial issue \cite{ERH}. The stumbling block for  some theorists using  this concept  is the absence of the spin-conservation law, which requires introducing source terms in the continuity equation for spin. There were worries about ambiguity of spin-current definition, and as an extreme of this stance, claims that the spin current has no physical meaning at all. Especially disturbing seemed the fact that  in  a 2D electron systems with the  Rashba spin-orbit term (let us call it {\em Rashba medium}) spin currents appear even in the equilibrium. Such currents were qualified as ``not real'', which cannot lead to transport and accumulation of spin \cite{R}. Their presence in the ground state was considered as an inherent problem in the spin current concept \cite{R1}.

It is worthwhile to note that the controversy about the spin-current concept is not new. The former round of the dispute on  spin transport was connected with discussions of possible dissipationless spin currents (spin supercurrents) in superfluid $^3$He and magnetically ordered solids (see the review Ref. \onlinecite{Usp} and references therein, or a more recent publication \cite{McD}).  At that time rejections of possibility of  dissipationless spin transport also appealed to nonconservation of spin. I would like to summarize here why ``fears'' of using the spin-current concept are not justified, or at least exaggerated:
\begin{itemize}
\item
There is no ban on using the concept of flow (flux, current) for nonconserving quantities and exploiting the continuity equations with source terms in physics. As an evident example I mention the balance of momentum in hydrodynamics. The continuity of the momentum contains the momentum-flux tensor and I am not aware that somebody worried whether this flux is real or not. 
\item
Appearance of currents (flows) in the equilibrium is not something unique for nonconserving spin. If a superconductor is in a magnetic field, equilibrium Meissner currents appear. These charge currents are circular and do not result in charge accumulation anywhere, but hardly somebody put in question their reality.

\item
Concerning ambiguity of spin current definition: Indeed, there is no sole definition of spin current, but this must not lead to any real ambiguity as well as no physical ambiguity arises from using different gauges in electrodynamics: various choices of potentials certainly (if correct) lead to equivalent physical predictions. Therefore numerous attempts to find a ``proper'' definition of spin current (see, e.g., Ref. \onlinecite{Niu}) do not seem sensible.
         \end{itemize}

In the present report I shall try to illustrate these statements considering equilibrium spin currents in the Rashba medium.

\section{Currents in the Rashba medium}

We consider a 2D electron gas with Rashba spin-obit interaction (Rashba medium). The essential physics is presented by the single-electron hamiltonian
\begin{eqnarray}
H ={\hbar^2\over 2 m}\left\{\vec \nabla \mathbf{\Psi}^\dagger \vec \nabla \mathbf{\Psi}
\right. \nonumber \\ \left.
 +i\alpha(\vec r) (\mathbf{\Psi}^\dagger  [\vec \sigma \times \hat z]_i \vec\nabla_i\mathbf{\Psi} -\vec \nabla_i \mathbf{\Psi}^\dagger[\vec \sigma \times \hat z]_i\mathbf{\Psi} )\right\}~,
              \end{eqnarray}
where  
 \begin{eqnarray}
\mathbf{\Psi} =\left( \begin{array}{c} \psi_\uparrow \\ \psi_\downarrow\end{array}\right)
              \end{eqnarray}
is a two-component spinor and $\vec \sigma$ is the vector of Pauli matrices.
In general the spin-orbit parameter $\alpha(\vec r)$ depends on the 2D position vector $\vec r$. The Schr\"odinger equations for components are:
 \begin{eqnarray}
i\hbar \dot \psi_\uparrow 
 ={\hbar^2 \over m}\left(-{1 \over 2}\nabla^2 \psi_\uparrow+\alpha {\partial \psi_\downarrow\over \partial x}-i \alpha {\partial \psi_\downarrow\over \partial y} 
 \right. \nonumber \\ \left.
+{1\over 2} {\partial \alpha\over \partial x}\psi_\downarrow-{i\over 2} {\partial \alpha\over \partial y}\psi_\downarrow\right)~,   \nonumber \\    
i\hbar\dot \psi_\downarrow 
 ={\hbar^2 \over m}\left(-{1 \over 2}\nabla^2 \psi_\downarrow-\alpha {\partial \psi_\uparrow\over \partial x}-i\alpha {\partial \psi_\uparrow\over \partial y}
 \right. \nonumber \\ \left.
  -{1\over 2} {\partial \alpha\over \partial x}\psi_\uparrow-{i\over 2} {\partial \alpha\over \partial y}\psi_\uparrow  \right)~.
           \end{eqnarray}           

The continuity equations for charge  and spin  can be derived either using Noether's theorem or directly from the Schr\"odinger equations multiplying them by complex-conjugated components: 
 \begin{eqnarray}
{\partial \rho \over \partial t}+\nabla_i J_i  =0 ~.
             \end{eqnarray} 
 \begin{eqnarray}
{\partial S_\beta \over \partial t}+\nabla_i J_i^\beta  =G_\alpha  ~.
          \label{SB}       \end{eqnarray} 
Here  $\rho=e|\mathbf{\Psi}|^2$ is the charge density,  $S_\alpha =(\hbar/2)( \mathbf{\Psi}^\dagger \sigma_\alpha\mathbf{\Psi})$ is the density of the $\alpha$ component of spin,
\begin{eqnarray}
J_i  =   -{ie\hbar \over 2m}( \mathbf{\Psi}^\dagger \nabla_i \mathbf{\Psi}-\nabla_i\mathbf{\Psi}^\dagger  \mathbf{\Psi}) 
- {\alpha e\hbar \over m} (\mathbf{\Psi}^\dagger [\vec \sigma \times \hat z]_i \mathbf{\Psi})   
             \end{eqnarray} 
is the charge current,
\begin{eqnarray}
J_i^\beta=-{i\hbar^2 \over 4 m}( \mathbf{\Psi}^\dagger \sigma_\beta \nabla_i \mathbf{\Psi} -\nabla_i\mathbf{\Psi}^\dagger \sigma_\beta  \mathbf{\Psi}) 
\nonumber \\
-{\alpha\hbar^2\over 4m} (\mathbf{\Psi}^\dagger \{\sigma_\beta [\vec \sigma \times \hat z]_i +[\vec \sigma \times \hat z]_i \sigma_\beta\} \mathbf{\Psi}) 
    \label{SC}         \end{eqnarray}
is the spin current. The Greek super(sub)script $\beta$ refers to three components $x,y,z$ in the 3D spin space, whereas the Latin super(sub)script $i$ is related with the two coordinates $x,y$ in the 2D electron layer.

The spin is not a conserved quantity, and there are source terms (torques) in the spin balance equations, 
 \begin{eqnarray}
G_\beta=-  {i\alpha\hbar^2\over 2m}\left\{  \left(\mathbf{\Psi}^\dagger\{ [\vec \sigma \times [\hat z \times \vec  \nabla]]_\beta \mathbf{\Psi}\} \right)
\right. \nonumber \\ \left.
 -\left(\{ [[\vec\nabla \times \hat z]\times \vec \sigma]_\beta \mathbf{\Psi}^\dagger\}   \mathbf{\Psi} \right)\right\}\,.
         \end{eqnarray}            

 Note that our choice of spin current,  Eq.~(\ref{SC}),  coincides with the current definition used by Rashba \cite{R} and many  others:
 \begin{eqnarray}
J^\beta_i ={\hbar \over 4}(\mathbf{\Psi}^\dagger\{ \sigma_\beta  v_i +v_i \sigma_\beta\}\mathbf{\Psi})\, ,
             \end{eqnarray}   
where 
\begin{eqnarray}
\vec v  ={\hbar\over m}(-i \vec \nabla+ \alpha  [\hat z \times   \vec  \sigma  ]) 
           \end{eqnarray}  
is the operator of the electron group velocity. This is a natural but not the {\em only} choice of current definition (see below). 

\section{Eigenstates and currents in an uniform Rashba medium}

\begin{figure}
  \begin{center}
    \leavevmode
    \includegraphics[width=0.9\linewidth]{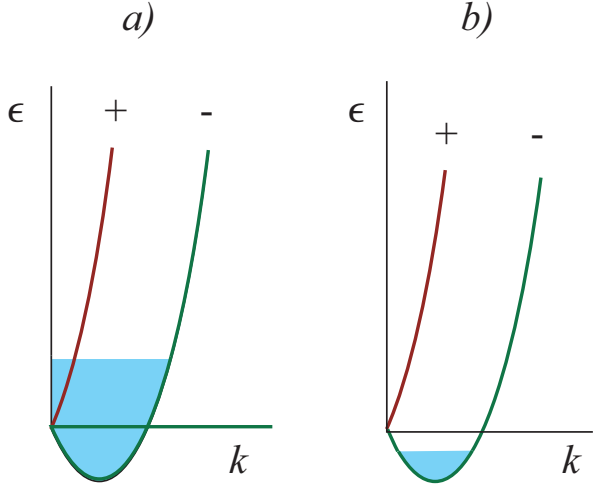}
    \caption{The ground state of the Rashba medium. a) The case $k_m >\alpha$, the Fermi sea (shaded blue) fills the upper (+) and the lower (-) band. b) The case $k_m<\alpha$, the Fermi sea fills only the lower band.}
  \label{fig1}
  \end{center}
  \end{figure}

    In the uniform Rashba medium eigenstates are plane waves given by spinors 
 \begin{eqnarray}
{1\over \sqrt{2}}\left( \begin{array}{c} 1 \\ \kappa \end{array}\right)e^{i\vec k  \vec r}~,
              \end{eqnarray}
where $\kappa = \pm(k_y-ik_x) /k$ are complex numbers of modulus 1, and the upper (lower) sign corresponds to the upper (lower) branch of the spectrum (band) with the energies (see Fig. \ref{fig1})      
 \begin{eqnarray}
\epsilon  ={\hbar^2\over m}\left({k^2\over 2}\pm \alpha k\right)={\hbar^2(k_0^2-\alpha ^2)\over 2 m}~.
           \label{ener}  \end{eqnarray}   
The energy is parametrized by the wave number  $k_0$, which is connected with absolute values of wave vectors in two bands as $k=|k_0 \mp \alpha|$.  
 The eigenstates are spin-polarized, and their spins are:
  \begin{eqnarray}
 \langle \vec  S \rangle=\pm {\hbar\over 2}{[\vec k \times \hat z] \over k}~.
             \end{eqnarray}   
The group velocities in two bands are given by
 \begin{eqnarray}
\vec v(\vec k)  ={\hbar \vec k\over m}+{2 \alpha \over m}  [\hat z \times  \langle \vec  S \rangle ] ={\hbar k_0\over m} {\vec k \over k}~.
         \label{gv}    \end{eqnarray}  
Spin torque in the eigenstates is absent, but  there are flows of spin components in the layer plane, i.e.,  spin currents, which {\em according to our definition of spin current} should be given by  
 \begin{eqnarray}
J^i_{j\pm}(\vec k) ={\hbar^2 \over 2m}\left(\pm {\varepsilon_{is} k_s \over k}k_j +\alpha \varepsilon_{ij}\right)~,
             \end{eqnarray}   
where $\varepsilon_{ij}$ is a 2D antisymmetric tensor with components $\varepsilon_{xy}=1$ and $\varepsilon_{yx}=-1$.

Though any eigenstate is spin-polarized,  after averaging over the equilibrium Fermi sea (we consider the $T=0$ case) the total spin vanishes. But there remain the total spin currents. The Fermi sea is determined by $k_m$, the maximum value of $k_0$, which determines  the Fermi energy $\epsilon_F=\hbar^2(k_m^2-\alpha^2)/2$. 
In the case $k_m>\alpha$, when the both electron bands are filled (Fig. \ref{fig1}a), the total electron density $n=\rho/e$ and spin currents are given by contributions from two bands: 
\begin{eqnarray} 
n=2\pi \int_0^{k_m-\alpha} k\,dk + 2\pi \int_0^{k_m+\alpha}k\,dk
=2 \pi(k_m^2+\alpha^2) ~.
            \end{eqnarray}  
\begin{eqnarray} 
J^i_j={\pi \hbar^2 \over m} \varepsilon_{ij}\left[\int_0^{k_m-\alpha} \left({k\over 2} +\alpha\right)k\,dk 
\right. \nonumber \\ \left.
+ \int_0^{k_m+\alpha}\left(-{k\over 2} +\alpha\right)k\,dk\right]  
={2\pi\alpha^3\hbar^2\over 3m}\varepsilon_{ij} ~.
            \end{eqnarray}   
At $k_m<\alpha$ only the lower band is filled (Fig. \ref{fig1}b), and
\begin{eqnarray} 
n=2\pi \int_{-k_m+\alpha}^{k_m+\alpha}k\,dk=4\pi \alpha k_m
=2 \pi(k_m^2+\alpha^2) ~.
            \end{eqnarray}  
\begin{eqnarray}
J^i_j= {\pi \hbar^2 \over m}\varepsilon_{ij} \int_{-k_m+\alpha}^{k_m+\alpha}\left(-{k\over 2} +\alpha\right)k\,dk
\nonumber \\
 ={\pi \hbar^2 \over m}\varepsilon_{ij} \left(-{k_m^3\over 3} + k_m\alpha^2\right)~.
             \end{eqnarray}

In contrast to charge currents, which produce magnetic fields, or mass currents in neutral superfluids, which produce angular momentum detectable in ring geometry  via the gyroscopic effect,  experimental detection of spin currents is much more difficult \cite{SCD}.  But this does not mean that equilibrium spin currents have no physical meaning and have nothing to do with spin transport. In order to demonstrate that spin current is able to transport spin we should consider monuniform media.

\section{spin currents in a nonuniform Rashba medium}

Let us consider a slightly modulated Rashba medium with the Rashba parameter varying in space as
 \begin{eqnarray} 
\alpha(\vec r)=\alpha_0+\alpha_1 \cos(\vec p\cdot\vec r) ~. 
             \end{eqnarray}   
The eigenstates found above must be corrected using the perturbation theory with respect to $\alpha_1$: $\mathbf{\Psi}=\mathbf{\Psi}_0+\mathbf{\Psi}'$. Here $\mathbf{\Psi}_0$ is the spinor for a uniform medium with $\alpha=\alpha_0$ with the components  
$\psi_{0\uparrow}=(1/\sqrt{2})e^{i\vec k\cdot \vec r}$ and  $\psi_{0\downarrow}=(\kappa /\sqrt{2})e^{i\vec k\cdot \vec r}$. The equations for the first order correction $\mathbf{\Psi}'$ are (in components):
 \begin{eqnarray}
{m\over \hbar^2}\Delta\epsilon\psi_\uparrow'- \alpha_0 [k \kappa^*(\vec k) +p \kappa^*(\vec p) ]\psi_\downarrow'
\nonumber \\
={\alpha_1 \kappa(\vec k)\over \sqrt{2}}\left[k \kappa^*(\vec k) +{p \kappa^*(\vec p)\over 2}\right] e^{i\vec k\cdot\vec r} \cos(\vec p\cdot\vec r)~,   
\nonumber \\    
-\alpha_0 [k \kappa(\vec k) +p \kappa(\vec p) ] \psi_\uparrow'+ {m\over \hbar^2}\Delta\epsilon\psi_\downarrow' 
\nonumber \\
={\alpha_1\over \sqrt{2}}\left[k \kappa^*(\vec k) +{p \kappa^*(\vec p)\over 2}\right] e^{i\vec k\cdot\vec r} \cos(\vec p\cdot\vec r)~,
            \end{eqnarray}     
where  
\begin{eqnarray}
\Delta\epsilon=-{\hbar^2\over m}\left({p^2 \over 2}+\vec p\cdot \vec k \mp \alpha_0 k\right) ~.
            \end{eqnarray}  
The solution of this system of linear equations for  $\mathbf{\Psi}'$ should be used for derivation of all relevant physical quantities (densities, torques, and currents). The general expressions are rather cumbersome. Moreover, the perturbation theory fails in the limit $p\to 0$. Therefore, it is reasonable to restrict ourselves with the limit $p \gg k, \alpha_0$.  The torques and currents for spin components in the plane of the layer are (only contributions linear in $\alpha_1$ are kept) 
\begin{eqnarray}
G_{i\pm}(\vec k) =\pm {2  \alpha_1\alpha_0\hbar^2\over m} \varepsilon_{ij}{p_jk\over p^2}\left[1-{(\vec p\cdot\vec k)^2\over p^2 k^2 }\right]\sin(\vec p\cdot\vec r)\,,
            \end{eqnarray}     
\begin{eqnarray}
J_{j\pm}^{i}(\vec k)= { \alpha_1\hbar^2\over 2 m}\left\{- p_j {  \varepsilon_{is}p_s\over p^2}+\varepsilon_{ij} 
\right. \nonumber \\ \left.
\mp{4\varepsilon_{ij} \alpha_0 k\over p^2} \left[1-{(\vec p\cdot\vec k)^2\over p^2k^2}\right]\right\} \cos (\vec p\cdot\vec r)\, .
            \end{eqnarray}    
The torque and the current for the $z$-component of spin are given by terms of higher order in $1/p$, and the total spin densities vanish after averaging over the Fermi sea. The integration of the torque and the current for inplane spin over the Fermi sea yields for the case $k_m >\alpha_0$:
\begin{eqnarray}
G_{i}= \int G_{i+}(\vec k)\,d\vec k + \int G_{i-}(\vec k)\,d\vec k
\nonumber \\=
 {2 \pi  \alpha_1\alpha_0\hbar^2\over m} {\varepsilon_{ij}p_j\over p^2}\left(\int_0^{k_m-\alpha_0}k^2dk
 \right. \nonumber \\ \left.
 -\int_0^{k_m+\alpha_0}k^2dk\right)\sin(\vec p\cdot\vec r)
 \nonumber \\
 =- {4 \pi  \alpha_1\alpha_0\hbar^2\over m} {\varepsilon_{ij}p_j\over p^2}\left(k_m^2\alpha_0+{\alpha_0^3\over 3}  \right)\sin(\vec p\cdot\vec r)~,
            \end{eqnarray}     
\begin{eqnarray}
J_{j}^{i}=\int J_{j+}^{i}(\vec k)\,d\vec k + \int J_{j-}^{i}(\vec k)\,d\vec k
\nonumber \\
= { \alpha_1\hbar^2\over 2 m}\left[ -\left(p_j {  \varepsilon_{is}p_s\over p^2}-\varepsilon_{ij}\right)n 
\right. \nonumber \\ \left.
+{8\pi\varepsilon_{ij} \alpha_0 \over p^2} \left(k_m^2\alpha_0+{\alpha_0^3\over 3}  \right)\right] \cos (\vec p\cdot\vec r)~.
            \end{eqnarray}    
If $k_m<\alpha_0$:
\begin{eqnarray}
G_{i}= \int G_{i-}(\vec k)\,d\vec k
\nonumber \\=
 {2 \pi  \alpha_1\alpha_0\hbar^2\over m} {\varepsilon_{ij}p_j\over p^2}\left(-\int_{-k_m+\alpha_0}^{k_m+\alpha_0}k^2dk\right)\sin(\vec p\cdot\vec r)
 \nonumber \\
 =- {4 \pi  \alpha_1\alpha_0\hbar^2\over m} {\varepsilon_{ij}p_j\over p^2}\left(k_m\alpha_0^2+{k_m^3\over 3}  \right)\sin(\vec p\cdot\vec r)~,
            \end{eqnarray}     
\begin{eqnarray}
J_{j}^{i}= \int J_{j-}^{i}(\vec k)\,d\vec k=
= { \alpha_1\hbar^2\over 2 m}\left[  -\left(p_j {  \varepsilon_{is}p_s\over p^2}-\varepsilon_{ij}\right)n 
\right. \nonumber \\ \left.
+{8\pi\varepsilon_{ij} \alpha_0 \over p^2}\left(k_m\alpha_0^2+{k_m^3\over 3}  \right)\right] \cos (\vec p\cdot\vec r)~.
            \end{eqnarray}  
The first term in the spin current, which is proportional to electron density $n$, is divergence-free, whereas the  divergence of the second term does not vanish and compensates the spin torque in  the spin balance. Thus the second term
is responsible for spin transport from areas, where spin is produced  ($G_i >0$) to areas where spin is absorbed ($G_i<0$). 

\section{Discussion and conclusions}

We have analyzed physical meaning of equilibrium spin currents in the Rashba medium  (a 2D electron gas with Rashba spin-orbit interaction). In a uniform Rashba medium these currents are constant and therefore do not lead to spin accumulation. Therefore some put in question  connection of these currents with {\em real} spin transport. But we have demonstrated that even weak inhomogeneity of the Rasha medium reveals ability of equilibrium spin currents to transport spin. Though in the case analyzed above spin accumulation is also absent, spin currents transfer spin from areas, where it is produced (as a result of spin-orbit interaction, which does not conserve spin) to areas, where spin is absorbed. This clearly demonstrates ability of equilibrium spin currents to transport spin. 

It is worthwhile also to comment again ambiguity of spin-current definition. Certainly one can redefine the spin current, which appears in the spin-balance equation (\ref{SB}), by adding to it {\em any} current $\tilde J_i^\beta$  ($J_i^\beta \to J_i^\beta +\tilde J_i^\beta$), if it is accompanied by redefinition of the spin torque: $G_\beta \to G_\beta +\nabla_i\tilde J_i^\beta$. This is a purely formal ambiguity of current definition (like freedom to choose various definitions of potentials in  electrodynamics), which must not lead to any ambiguity in physical predictions. This  only means that any definition of current is not complete without accompanying definition of spin torque. It is important also that no choice of current definition can eliminate spin torque completely, since the latter cannot be reduced to a divergence of some current if spin is not conserved indeed.

Arguing  full consistency of the spin current concept despite non-conservation of spin, we do not mean an {\em obligation} to use this concept. This is nothing more than one of possible languages for description of spin processes. It is possible to use other descriptions for the same phenomena.  For example, one may treat the whole current divergence $\nabla_i J_i^\beta $ as a part of the spin torque.  We already stressed similarity of the spin current concept  with the concept of momentum flux in hydrodynamics. In the elasticity theory they use the stress tensor instead of the momentum-flux tensor, but the difference is only in semantics. Describing spin processes one can also avoid such ``dangerous'' (as some people believe) terms as flow or current, and describe the same phenomena using only concepts of deformation, rigidity, or torque (as discussed  in  Ref. \onlinecite{Usp} with respect to magnetically ordered systems). However, in many cases the ``current language'' provides a very transparent physical picture of processes in various spin systems.

\section*{Acknowledgment}

The work was supported by the grant of the Israel Academy of
Sciences and Humanities.



\begin{thebibliography}{99}
\bibitem{ZFS} I. {\u Z}uti{\' c}, J. Fabian, and S. Das Sarma, Rev. Mod. Phys. {\bf 76}, 323 (2004).
\bibitem{ERH} H.-A. Engel, E.I. Rashba, and B.I. Halperin, cond-mat/0603306.
\bibitem{R} E.I. Rashba,  Phys. Rev. B {\bf 68}, 241315 (2003).
\bibitem{R1} E.I. Rashba,  J. Supercond.  {\bf 18}, 1137 (2005); Physica E {\bf 34}, 31 (2006).
\bibitem{Usp} E.~B. Sonin,  Usp. Fiz. Nauk {\bf137}, 267 (1982) [Sov. Phys.--Usp. {\bf 25}, 409 (1982)].
\bibitem{McD} J. K\"onig, M. C. B{\o}nsager, and A.H. MacDonald, Phys. Rev. Lett. {\bf 87}, 187202 (2001).
\bibitem{Niu} J. Shi, P. Zhang, D. Xiao, and Q. Niu, Phys. Rev. Lett. {\bf 96}, 076604 (2006).
\bibitem{SCD} Spin current leads to electric polarization (magnetoelectric effect). This can be used for current detection as discussed by H. Katsura, N. Nagaosa, and A.V. Balatsky, Phys. Rev. Lett. {\bf 95}, 057205 (2005), and by  P. Bruno and V.K. Dugaev, Phys. Rev. B {\bf 72}, 241302(R) (2005).
\end{thebibliography}
\end{document}